\documentclass[12pt]{article}
\usepackage[letterpaper,margin=1in]{geometry}
\usepackage[T1]{fontenc}
\usepackage[utf8]{inputenc}
\usepackage{lmodern}
\usepackage{amsmath,amssymb,bm,mathtools}
\usepackage{booktabs,array}
\usepackage{microtype}
\usepackage{enumitem}
\usepackage{xurl}
\usepackage{xcolor}
\usepackage[hypertexnames=false,colorlinks=true,linkcolor=blue!45!black,citecolor=blue!45!black,urlcolor=blue!55!black]{hyperref}

\newcommand{\Tr}{\operatorname{Tr}}
\newcommand{\Cov}{\operatorname{Cov}}

\newcommand{\ii}{\mathrm{i}}
\newcommand{\Hilbert}{\mathcal{H}}

\title{\textbf{Spin Reactions in Density Matrix Form}\\[0.75em]
\large From the Chiang--Tabakin complete-experiment analysis to threshold-dependent reduced descriptions}
\author{Yannick Wunderlich$^{1}$ and Frank Tabakin$^{2}$\\[0.5em]
\normalsize $^{1}$Independent Researcher, D-50321 Brühl, Germany\\
\normalsize $^{2}$Department of Physics and Astronomy, University of Pittsburgh,\\
\normalsize Pittsburgh, Pennsylvania 15260, USA}
\date{November 2026}

\begin{document}
\maketitle

\begin{abstract}
The complete-experiment problem asks which polarization measurements determine a reaction amplitude up to its physically irrelevant overall phase. For pseudoscalar-meson photoproduction, the Chiang--Tabakin analysis organized the sixteen observables as bilinear forms in four transversity amplitudes and used the completeness algebra of $4\times4$ matrices and associated Fierz identities to derive observable constraints and complete sets. We present an equivalent formulation in terms of a normalized rank-one density matrix in amplitude space. In this language, the Fierz identities are precisely the component form of density-matrix purity. Reduced density matrices and their submatrices expose which amplitude populations, coherences, and correlations survive when polarization information is discarded. The formalism is applied first to pseudoscalar-meson photoproduction and then to pion--nucleon scattering, where two amplitudes yield the familiar unit-sphere relation $P^2+A^2+R^2=1$. Finally, we distinguish tensor-product partial traces from traces over direct-sum reaction channels.
The amplitude space density matrix( ADM) discussed here is not a spin-space density matrix.  When additional channels open at threshold, an exclusive channel ADM retains its pure-state identities, whereas an unresolved inclusive measurement generally produces a mixed reduced density matrix and converts purity equalities into positivity inequalities. This provides a unified interpretation of Fierz constraints, subchannels, submatrices and partial traces.
The present manuscript traces the aforementioned concepts and methods back to a single object - the ADM density matrix - and its fundamental quantum mechanical postulates. This reduction and synthesis appears in a few papers, but deserves to be expanded and developed as a tool for formulating and constraining general reaction amplitudes.
\end{abstract}

\section{Historical background}
Polarization experiments were recognized early as a route from measured cross sections to complex reaction amplitudes (cf.~\cite{Barker1975,Fasano1992} as well as summaries on reaction formalisms and experimental results given in~\cite{Thiel2022}). In pseudoscalar-meson photoproduction, parity reduces the process to four independent complex amplitudes. The central difficulty is that observables are quadratic in these amplitudes: an overall phase is unobservable, and incomplete bilinear information leaves discrete ambiguities. Barker, Donnachie, and Storrow formulated influential complete-experiment criteria in 1975~\cite{Barker1975}. Subsequent work clarified ambiguity structures and the role of different amplitude bases.

Chiang and Tabakin supplied a systematic algebraic solution in 1997~\cite{Chiang1997}. They represented each observable by a Hermitian $4\times4$ matrix acting on the four-amplitude vector. The orthogonality and completeness of this matrix basis permit Fierz rearrangements, yielding linear and nonlinear relations among the sixteen observables. They showed that eight carefully selected measurements can determine the four transversity amplitudes without discrete ambiguity, apart from the unavoidable common phase. This sharpened the earlier counting and made the observable-class structure operational.

Later studies~\cite{Wunderlich2014,Nakayama2019,Wunderlich2020,Wunderlich2021} revisited complete sets in the presence of experimental uncertainty, truncated partial-wave analyses, and exceptional phase configurations. Wunderlich and collaborators~\cite{Wunderlich2020,Wunderlich2021} connected the problem to graph-theoretical forms of Moravcsik's theorem and treated $\pi N$ scattering ($N=2$), photoproduction ($N=4$), and electroproduction ($N=6$) in a common language. The present paper adds a density-matrix formulation designed to make the physical content of Fierz relations, reductions, and channel opening especially transparent. Furthermore, it will be shown that using this density-matrix formulation, the intrinsic connections between Fierz relations, the Chiang--Tabakin method, and graph-theoretic methods will become particularly evident.

\section{Bilinear observables and the Chiang--Tabakin algebra}
Let $\bm b=(b_1,\ldots,b_N)^{\mathsf T}$ be a column of complex amplitudes. A complete orthogonal Hermitian matrix basis $\Gamma^{(\alpha)}$ satisfies
\begin{equation}
 \Tr\!\left[\Gamma^{(\alpha)}\Gamma^{(\beta)}\right]
 =N\delta_{\alpha\beta}.
 \tag{1}
\end{equation}
Profile functions---observables multiplied by the unpolarized intensity---are bilinear forms
\begin{equation}
 \widehat O_\alpha=c_\alpha\,\bm b^\dagger\Gamma^{(\alpha)}\bm b,
 \tag{2}
\end{equation}
where $c_\alpha$ carries normalization and sign conventions. Completeness of the $\Gamma$ basis permits a rearrangement of products of bilinears. The resulting Fierz identities~\cite{Chiang1997} are not extra dynamical assumptions. They follow because all observables are constructed from the same finite set of amplitudes.

For $N=4$ photoproduction, the sixteen observables separate into the unpolarized/single-spin class $S$ and three double-polarization classes: beam--target (BT), beam--recoil (BR), and target--recoil (TR). Different sign conventions are common, but the class structure and invariant content are unchanged.
\begin{center}
\begin{tabular}{@{}ll@{}}
\toprule
Class & Representative observables\\
\midrule
$S$  & $\sigma_0,\ \Sigma,\ T,\ P$\\
BT   & $E,\ F,\ G,\ H$\\
BR   & $C_{x'},\ C_{z'},\ O_{x'},\ O_{z'}$\\
TR   & $T_{x'},\ T_{z'},\ L_{x'},\ L_{z'}$\\
\bottomrule
\end{tabular}
\end{center}

\section{Equivalent amplitude density-matrix (ADM) formulation}
Define the normalized amplitude density matrix (ADM)
\begin{equation}
 \rho=\frac{|b\rangle\langle b|}{\langle b|b\rangle},
 \ \ \ \  \Tr\rho=1.
 \tag{3}
\end{equation}
At fixed kinematics and for one coherent exclusive amplitude, $\rho$ has rank one. Therefore
\begin{equation}
 \rho^2=\rho,\ \ \   \Tr\rho^2=1,\ \ \   \det\rho=0.
 \tag{4}
\end{equation}
Expand $\rho$ in the same orthogonal matrix basis used for the observables:
\begin{equation}
 \rho=\frac{1}{N}\sum_\alpha r_\alpha\Gamma^{(\alpha)},
 \ \ \  r_\alpha=\Tr\!\left[\rho\Gamma^{(\alpha)}\right].
 \tag{5}
\end{equation}
The coefficients $r_\alpha$ are normalized observables, subject to convention-dependent relabelings and signs. The purity constraint for the matrix $\rho$ is most generally formulated in terms of the vanishing of all of its $2\times2$ minors, and it yields exactly the same quadratic relations produced by the Fierz rearrangement. This fact is proved in detail in Appendix~A. Thus Fierz identities and density-matrix purity are two representations of one algebraic fact: the $N^2$ observable components are generated by a single ray in an $N$-dimensional complex amplitude space.

This equivalence also makes counting transparent. A normalized pure state in $\mathbb C^N$ contains $2N-2$ real parameters; restoring the overall intensity adds one. More observables may be required experimentally to eliminate discrete ambiguities, but the identities ensure that the $N^2$ measured quantities are highly constrained.

The Chiang--Tabakin procedure can be formulated and proven using the Fierz identities~\cite{Chiang1997} and it is thus deeply rooted in density-matrix purity constraints. Furthermore, the seemingly different methods for deriving complete sets using graph-theoretical ideas as in references~\cite{Wunderlich2020,Wunderlich2021} are also in principle a result of density-matrix purity. This is true since density-matrix purity produces the consistency relations among amplitude relative phases $\phi_{ij}:=\phi_i-\phi_j$ (with $b_j=|b_j|e^{i\phi_j}$), which lie at the heart of such graphical construction methods. More details and a proof of this connection are provided in Appendix~B.

\subsection{A decision problem rather than a predictive calculation}
The Chiang--Tabakin construction should be understood primarily as a decision and identifiability problem. Given a menu of experimentally measurable observables, it decides which subsets are algebraically sufficient to reconstruct the amplitudes, up to the unavoidable overall phase, and which selections leave discrete ambiguities. It does not predict the numerical values of amplitudes or observables. Such predictions require a dynamical model, a partial-wave analysis, or a fit to data. The C--T analysis instead specifies what must be measured so that competing dynamical descriptions can, in principle, be distinguished. \footnote{ Quantum Metrology is the term used to describe how to extract system parameters from quantum system measurements, see arXiv:2608.01617v1[quant-p] 3 Aug 2026.}

Algebraic completeness assumes exact observables. Real measurements have error bars, correlated normalization uncertainties, finite acceptance, and systematic effects. These uncertainties can turn a formally complete set into a practically weak one: two different amplitude solutions may remain statistically indistinguishable even though exact data would separate them. Consequently, realistic completeness must combine the C--T ambiguity test with the covariance matrix of the measured observables.

\subsection{Fisher information and realistic information content}
Let $\theta_a$ denote independent amplitude parameters---for example, moduli and relative phases after fixing the overall phase---and let $\bm O(\bm\theta)$ be the vector of predicted observables. For approximately Gaussian measurement errors with covariance matrix $V$, the local Fisher information matrix is
\begin{equation}
 F_{ab}=\left(\frac{\partial\bm O}{\partial\theta_a}\right)^{\!\mathsf T}
 V^{-1}\left(\frac{\partial\bm O}{\partial\theta_b}\right).
 \tag{5a}
\end{equation}
The error bars enter through $V$: an observable with a large uncertainty contributes less information, while correlated systematic errors prevent nominally separate measurements from being counted as independent. Large Fisher eigenvalues identify well-measured parameter combinations; small eigenvalues expose weak or nearly degenerate directions. Subject to the usual regularity and local-Gaussian assumptions, the inverse matrix gives the Cram\'er--Rao covariance bound
\begin{equation}
 \Cov(\widehat{\bm\theta})\succeq F^{-1}.
 \tag{5b}
\end{equation}
Fisher information therefore supplies a realistic measure of local information content and a natural criterion for choosing among algebraically complete sets. D-optimal choices maximize $\det F$ and reduce the total uncertainty volume; E-optimal choices maximize the smallest eigenvalue and protect the weakest direction. Fisher analysis is local, however, and does not by itself detect disconnected discrete solutions. A sound design uses both layers: Chiang--Tabakin/Fierz analysis for global algebraic identifiability, followed by likelihood or Fisher analysis with full error propagation for achievable precision~\cite{Fisher1922,Rao1945}.

\section{Submatrices, reduced matrices, and partial traces}
\subsection{Principal submatrices}
A principal submatrix is obtained by selecting a subset $S$ of amplitude indices: $\rho_S=P_S\rho P_S$, where $P_S$ is the corresponding projector. Its diagonal entries are relative populations $|b_i|^2/I_0$; its off-diagonal entries are coherences $b_i b_j^*/I_0$. Positivity of every principal submatrix implies
\begin{equation}
 \rho_{ii}\geq0,\qquad |\rho_{ij}|^2\leq\rho_{ii}\rho_{jj}.
 \tag{6}
\end{equation}
For a rank-one pure matrix every two-index principal minor vanishes, so $|\rho_{ij}|^2=\rho_{ii}\rho_{jj}$ for every pair. In a reduced or mixed description the inequalities need not saturate. These minors therefore supply directly testable positivity conditions on observable combinations.

\subsection{Partial traces in a Pauli-product organization}
For $N=4$ one may organize amplitude space algebraically as $\mathbb C^4\cong\mathbb C^2\otimes\mathbb C^2$ and expand
\begin{equation}
 \rho=\frac14\sum_{\mu,\nu}C_{\mu\nu}\,\sigma_\mu\otimes\sigma_\nu.
 \tag{7}
\end{equation}
The partial traces are
\begin{equation}
 \rho_1=\Tr_2\rho=\frac12\sum_\mu C_{\mu0}\sigma_\mu,
 \ \ \ 
 \rho_2=\Tr_1\rho=\frac12\sum_\nu C_{0\nu}\sigma_\nu.
 \tag{8}
\end{equation}
They retain single-factor polarization information and remove the correlation coefficients $C_{ij}$. Physically, a partial trace represents summing over a polarization label that is not resolved. However, the two factors in Eq.~(7) are initially an algebraic factorization of amplitude space, not automatically two particles. A beam, target, or recoil interpretation is valid only after the basis-to-experiment mapping has been specified.

\section{Application to pseudoscalar-meson photoproduction}
For $\gamma N\to MN'$ with a pseudoscalar meson $M$, four complex amplitudes generate sixteen observables. In the transversity basis, the four $S$-class quantities determine the four moduli $|b_i|$, subject to normalization and positivity. The BT, BR, and TR classes contain real and imaginary parts of interference products and hence constrain relative phases. A complete experiment combines modulus information with enough cross-class interference information to connect all amplitudes and eliminate discrete phase choices.

The density matrix provides an immediate diagnostic hierarchy. First, $\rho$ must be Hermitian, positive semidefinite, and normalized. Second, for an exclusive coherent reaction at a fixed kinematic point it must satisfy $\Tr\rho^2=1$. Third, all Fierz constraints follow from the same rank-one condition. Experimental central values need not satisfy every equality exactly because of statistical fluctuations, acceptance corrections, finite bins, and inconsistent convention choices. A useful analysis can therefore fit the nearest physical density matrix rather than enforce individual identities one at a time.

The reduced matrices in Eq.~(8) identify which combinations remain accessible when one polarization sector is ignored. Their Bloch-vector lengths obey $|\bm r_1|\leq1$ and $|\bm r_2|\leq1$. Equality means that the retained effective sector is itself pure; a strict inequality signals correlations with the traced factor or incoherent averaging. In this sense, the submatrices are not merely bookkeeping devices: their eigenvalues quantify how much polarization information remains locally and how much resides in correlations.

\subsection{Experimental uses of the observable constraints}
The impact of the Chiang--Tabakin program is especially visible in hyperon photoproduction, where the self-analyzing weak decay of the $\Lambda$ makes recoil polarization experimentally accessible. The CLAS measurement by Bradford et al. reported the first beam--recoil observables $C_x$ and $C_z$ for $\gamma p\to K^+\Lambda$ and $\gamma p\to K^+\Sigma^0$. For $K^+\Lambda$, the magnitude of the recoil-polarization vector formed from $P$, $C_x$, and $C_z$ was found to be consistent with unity over much of the measured region. Although this result is not by itself a complete-experiment reconstruction, it is a striking empirical realization of the quadratic spin-length constraints that follow from the bilinear formalism~\cite{Bradford2007}.

Paterson et al. subsequently measured $\Sigma$, $T$, $O_x$, and $O_z$ for $K^+\Lambda$ and $K^+\Sigma^0$ photoproduction with a linearly polarized photon beam. Together with earlier $P$, $C_x$, and $C_z$ measurements, these observables populate several Chiang--Tabakin classes and strongly constrain amplitude analyses. The publication explicitly used a Fierz relation connecting $O_x$, $O_z$, $C_x$, $C_z$, $\Sigma$, $T$, and $P$ as an internal consistency test and as a check on polarization calibration~\cite{Paterson2016}. This illustrates an important practical role of Fierz identities: they test whether independently extracted asymmetries can arise from one physical amplitude matrix.

A particularly direct application was given by Ireland et al. They combined extensive kaon-photoproduction polarization data and used Fierz constraints to determine calibration combinations involving the photon polarization and the $\Lambda$ weak-decay parameter $\alpha_-$. Their result, $\alpha_-=0.721(6)(5)$, supported the then-new BESIII determination and demonstrated that the Chiang--Tabakin identities can constrain an external calibration parameter, not merely test an amplitude fit~\cite{Ireland2019}. These examples show the progression from measuring individual observable classes, through testing their algebraic compatibility, to using the constraints quantitatively in a global experimental analysis.

In all such applications, error bars are part of the physics conclusion rather than decoration on an exact identity. A residual from a Fierz relation should be compared with its propagated uncertainty, including covariance among observables that share events, beam-polarization calibration, target polarization, acceptance, or the same recoil-decay parameter. The statistically meaningful question is whether the residual is significant relative to that joint uncertainty. Likewise, the information supplied by an added observable depends on its precision and on whether it probes a new derivative direction in amplitude space; the Fisher matrix makes this distinction quantitative.

\section{Application to pion--nucleon scattering}
Elastic scattering of a spin-zero pion from a spin-$\tfrac12$ nucleon requires two complex amplitudes. Write $\bm\psi=(f,g)^{\mathsf T}$ and $I_0=|f|^2+|g|^2$. One convenient convention defines
\begin{equation}
 A=\frac{|f|^2-|g|^2}{I_0},\ \ \ 
 R=\frac{2\operatorname{Re}(f^*g)}{I_0},\ \ \ 
 P=\frac{2\operatorname{Im}(f^*g)}{I_0}.
 \tag{9}
\end{equation}
The normalized density matrix (ADM) is
\begin{equation}
 \rho_{\pi N}=\frac12\left(I+R\sigma_x+P\sigma_y+A\sigma_z\right).
 \tag{10}
\end{equation}
Purity, $\det\rho_{\pi N}=0$, or the two-dimensional Pauli Fierz identity all give
\begin{equation}
 P^2+A^2+R^2=1.
 \tag{11}
\end{equation}
This is the complete geometric content of the two-amplitude exclusive problem: the three normalized observables lie on the Bloch sphere. Reversing the normal to the scattering plane changes convention-dependent signs but not Eq.~(11). If unobserved processes, finite-bin averaging, or unresolved channels produce a mixed effective state, the result becomes the positivity bound $P^2+A^2+R^2\leq1$.

\subsection{Worked two-channel example}
To make the channel reduction explicit, consider a stipulated illustrative point at total center-of-mass energy $W=1.30\,\mathrm{GeV}$. Take the elastic $\pi N$ channel and the effective $\pi\pi N$ channel to be open, with intensity fractions $w_1=0.70$ and $w_2=0.30$. Take $\eta N$ and $K\Lambda$ to be closed. These numbers are chosen to expose the algebra and are not a fit to $\pi N$ data.

Choose normalized two-amplitude spinors for the open channels:
\begin{equation}
 |\chi_1\rangle=
 \begin{pmatrix}\sqrt3/2\\[2pt]1/2\end{pmatrix},
 \ \ \ 
 |\chi_2\rangle=
 \begin{pmatrix}1/\sqrt2\\[2pt]\ii/\sqrt2 \end{pmatrix}.
 \tag{11a}
\end{equation}
With the convention of Eq.~(9), channel 1 has $(A_1,R_1,P_1)=(1/2,\sqrt3/2,0)$, while channel 2 has $(A_2,R_2,P_2)=(0,0,1)$. Thus each resolved exclusive channel lies on its own Bloch sphere:
\begin{equation}
 A_1^2+R_1^2+P_1^2=1,\qquad
 A_2^2+R_2^2+P_2^2=1.
 \tag{11b}
\end{equation}
The corresponding normalized exclusive density matrices are
\begin{equation}
 \bar\rho_1=
 \begin{pmatrix}3/4&\sqrt3/4\\[2pt]\sqrt3/4&1/4\end{pmatrix},
 \ \ \ 
 \bar\rho_2=
 \begin{pmatrix}1/2&-\ii/2\\[2pt]\ii/2&1/2\end{pmatrix}.
 \tag{11c}
\end{equation}
The channel projectors select two orthogonal direct-sum sectors. Their unnormalized surviving subchannel matrices are $w_1\bar\rho_1$ and $w_2\bar\rho_2$:
\begin{equation}
 \rho_1^{(\mathrm{un})}=
 \begin{pmatrix}0.5250&0.3031\\0.3031&0.1750\end{pmatrix},
 \ \ \ 
 \rho_2^{(\mathrm{un})}=
 \begin{pmatrix}0.1500&-0.1500\ii\\0.1500\ii&0.1500\end{pmatrix}.
 \tag{11d}
\end{equation}
If the final channel is identified, the appropriate matrix is $\bar\rho_1$ or $\bar\rho_2$ and the pure-state equality remains exact. If the detector accepts both channels but does not retain the channel label, the channel subtrace is the incoherent sum
\begin{align}
 \bar\rho_{\mathrm{spin}}
 &=\Tr_{\mathrm{channel}}\rho_{\mathrm{out}}
 =0.70\bar\rho_1+0.30\bar\rho_2 \notag\\
 &=\begin{pmatrix}
 0.6750&0.3031-0.1500\ii\\
 0.3031+0.1500\ii&0.3250
 \end{pmatrix}.
 \tag{11e}
\end{align}
The measured inclusive Bloch components are therefore the intensity-weighted averages
\begin{equation}
 (A,R,P)=(0.3500,0.6062,0.3000),\qquad
 A^2+R^2+P^2=0.5800<1.
 \tag{11f}
\end{equation}
The two eigenvalues and the purity are
\begin{equation}
 \lambda_\pm=\frac{1\pm\sqrt{0.58}}{2}=(0.8808,0.1192),
 \quad \Tr(\bar\rho_{\mathrm{spin}}^2)=0.7900,
 \quad \det\bar\rho_{\mathrm{spin}}=0.1050.
 \tag{11g}
\end{equation}
This strict inequality is not a failure of the exclusive $\pi N$ identity. It records information lost by discarding which open channel occurred. The off-diagonal channel coherences vanish because the $\pi N$ and $\pi\pi N$ final states are orthogonal; only their positive density-matrix blocks survive. In this example, the subchannel matrices display what each projector retains, while the channel subtrace displays what an unresolved measurement reports.

\section{Thresholds, additional channels, and channel traces}
The opening of an additional physical channel must be distinguished from the algebraic factorization used in Eq.~(7). Reaction channels are orthogonal sectors of a direct sum,
\begin{equation}
 \Hilbert_{\mathrm{out}}=\bigoplus_c\Hilbert_c.
 \tag{12}
\end{equation}
For an initial state $|i\rangle$, the full outgoing state may be written
\begin{equation}
 |\Psi_{\mathrm{out}}\rangle=\sum_c|c\rangle\otimes|\psi_c\rangle,
 \tag{13}
\end{equation}
where $|c\rangle$ labels the physical channel and $|\psi_c\rangle$ contains its spin amplitudes~\footnote{ Note $|\psi_c\rangle= (1+ G_0 T) \Phi,$  where the spin matrix elements of $T$ are the spin amplitudes for an incident state  $\Phi$. }. Below the threshold $E_c^{(\mathrm{th})}$, channel $c$ has no on-shell phase space, although virtual effects can modify amplitudes analytically. Above threshold its momentum becomes real, its phase-space factor turns on, and unitarity acquires a new absorptive contribution. This produces familiar cusps, rapid phase motion, and redistribution of flux among exclusive reactions.

If the experiment resolves $c$, one constructs the exclusive density matrix
\begin{equation}
 \rho_c=\frac{|\psi_c\rangle\langle\psi_c|}{\langle\psi_c|\psi_c\rangle}.
 \tag{14}
\end{equation}
At fixed kinematics this matrix is again pure, so the channel-specific Fierz equalities remain valid. If the channel label is unobserved, one traces it out:
\begin{equation}
 \rho_{\mathrm{spin}}=\Tr_{\mathrm{channel}}
 |\Psi_{\mathrm{out}}\rangle\langle\Psi_{\mathrm{out}}|
 =\sum_c|\psi_c\rangle\langle\psi_c|.
 \tag{15}
\end{equation}
After normalization, Eq.~(15) is generally mixed. Interference between orthogonal, experimentally unresolved channel labels disappears under the trace; their probabilities add. The observable constraints therefore weaken from rank-one equalities to positive-semidefinite inequalities. This is the precise relation between subchannels and subtraces: a subchannel is a projected physical sector $P_c\Hilbert_{\mathrm{out}}$, whereas the channel partial trace is the operation that discards which-sector information and adds the surviving sector density matrices incoherently.

\section{Observable constraints across a threshold}
Let $w_c=\Tr\rho_c^{(\mathrm{un})}$ be the intensity in channel $c$ and let $\bar\rho_c$ be its normalized spin density matrix. An unresolved sample has
\begin{equation}
 \bar\rho=\frac{\sum_c w_c\bar\rho_c}{\sum_c w_c}.
 \tag{16}
\end{equation}
Its purity is bounded by $\Tr\bar\rho^2\leq1$. Equality holds only when a single channel contributes or all contributing channels have the same normalized spin state. Consequently, deviations from exclusive Fierz equalities can have several origins: ordinary experimental uncertainty, kinematic averaging, incomplete polarization preparation, or genuine unresolved-channel mixing. A threshold study must separate these effects before interpreting a reduced purity as new dynamics.

For $\pi N$ observables the inclusive relation becomes $P^2+A^2+R^2\leq1$. For four-amplitude photoproduction, the analogous statement is that the reconstructed $4\times4$ matrix must remain positive semidefinite while its rank can exceed one. Principal-minor inequalities and $\Tr\rho^2$ provide compact diagnostics. Exclusive measurements of each open channel should recover the appropriate pure-state identities independently.

\section{Discussion and outlook}
The density-matrix language does not replace the Chiang--Tabakin Fierz analysis; it reveals its geometry. The original $\Gamma$-matrix completeness algebra produces relations among observable components. The alternative formulation assembles those components into a positive operator and identifies the relations with normalization, positivity, and purity. Principal submatrices isolate populations and coherences. Partial traces specify which information has been deliberately discarded. Direct-sum channel projectors identify physical reaction sectors, while a trace over channel labels describes an unresolved inclusive measurement.

This structure suggests practical extensions. Experimental databases can be tested by reconstructing $\rho$ and monitoring its eigenvalues and purity. Fits can impose positivity globally instead of checking isolated inequalities. Threshold scans can compare channel-resolved and inclusive reconstructions, distinguishing kinematic mixing from changes in the underlying spin amplitudes. The same framework extends naturally from $\pi N$ scattering and pseudoscalar photoproduction to electroproduction and reactions with larger amplitude spaces.

\section{Conclusions}
Fierz identities and density-matrix purity are equivalent expressions of amplitude completeness. In pseudoscalar-meson photoproduction, four amplitudes produce sixteen constrained observables organized into $S$, BT, BR, and TR classes. In $\pi N$ scattering, the corresponding two-amplitude geometry reduces to $P^2+A^2+R^2=1$. Submatrices and partial traces have clear but distinct meanings: submatrices select amplitude sectors and expose positivity; polarization partial traces discard specified spin information; channel partial traces discard which-channel information. When a new physical channel opens, each resolved exclusive channel retains its pure-state identities, while an unresolved sum generally becomes mixed and satisfies inequalities rather than equalities. This distinction supplies the desired connection among observable constraints, subchannels, and subtraces.

Excellent papers~\cite{Durupt2026,Artru2009} using ADM descriptions are available, but do not establish the equivalence to using Fierz identities nor make application to these specific reactions.  Our hope is to expand this work to include the effects of opening channels and also to the energy dependence of the reaction amplitudes.  For example. the Wigner-Smith operator, incorporates limited causality conditions based on wave packet analysis. A full treatment,  starting from coupled channels Lippmann-Schwinger and Faddeev equations, cast into an ADM description and possibly Lindbladian-type energy evolution would be instructive.  A fully covariant treatment with causal propagators would be an additional advance.

\section*{Acknowledgements}

The authors used ChatGPT (OpenAI, GPT-6.1 Sol, accessed October  2026) to assist in developing explanations of the amplitude-density-matrix formalism and preparing mathematical text and LaTeX. The authors directed this assistance, independently checked the derivations and references, and revised the resulting material. The authors take full responsibility for the scientific content and conclusions.

\appendix
\section{Equivalence of purity constraints for the density matrix $\rho$ and Fierz relations}
In this appendix, we prove the claimed equivalence between the purity of the density matrix $\rho$, expressed as $\operatorname{Rank}(\rho)=1$, and the Fierz identities satisfied by polarization observables. For simplicity, the proof is written in terms of profile functions $\widehat O_\alpha$ and the non-normalized density matrix $R:=\langle b|b\rangle\rho$. The proof can also be rewritten in terms of the normalized quantities defined in the main text.

\paragraph{$\operatorname{Rank}(R)=1\Rightarrow$ Fierz.}
We write profile functions as (repeated indices summed, and normalization factors $c_\alpha$ dropped for simplicity):
\begin{equation}
\widehat O_\alpha=b_j^*\Gamma^{(\alpha)}_{ji}b_i
=R_{ij}\Gamma^{(\alpha)}_{ji}=\Tr[R\Gamma^{(\alpha)}].\tag{A.1}
\end{equation}
For the positive-semidefinite Hermitian matrix $R=bb^\dagger$, the rank-one property is equivalent to the vanishing of all $2\times2$ minors,
\begin{equation}R_{ij}R_{kl}=R_{il}R_{kj}.\tag{A.2}\end{equation}
Combining this with completeness of the $\Gamma$ matrices, $R=N^{-1}\sum_\gamma\widehat O_\gamma\Gamma^{(\gamma)}$, gives
\begin{align}
\widehat O_\alpha\widehat O_\beta
&=R_{ij}R_{kl}\Gamma^{(\alpha)}_{ji}\Gamma^{(\beta)}_{lk}\notag\\
&=R_{il}R_{kj}\Gamma^{(\alpha)}_{ji}\Gamma^{(\beta)}_{lk}
=\Tr[R\Gamma^{(\alpha)}R\Gamma^{(\beta)}]\notag\\
&=\sum_{\gamma,\delta}\underbrace{\frac{1}{N^2}\Tr[\Gamma^{(\gamma)}\Gamma^{(\alpha)}\Gamma^{(\delta)}\Gamma^{(\beta)}]}_{C^{\alpha\beta}_{\gamma\delta}}
\widehat O_\gamma\widehat O_\delta.\tag{A.3}
\end{align}
This agrees with the Fierz identities in reference~\cite{Chiang1997} and illustrates that
\begin{equation}2\times2\text{ minors}+\Gamma\text{-completeness}\ \Longrightarrow\ \text{Fierz relations}.\tag{A.4}\end{equation}
The same local rank-one constraints generate graphical cycle-consistency relations. Repeated application of the minor identities reduces arbitrary cycle products to products of diagonal elements; see Appendix~B.

\paragraph{Fierz $\Rightarrow\operatorname{Rank}(R)=1$.}
Define the tensor
\begin{equation}M_{ij;kl}:=R_{ij}R_{kl}-R_{il}R_{kj}.\tag{A.5}\end{equation}
Using the unique expansion in the orthonormal $\Gamma$ basis,
\begin{equation}
M_{ij;kl}=\frac1{N^2}\sum_{\alpha,\beta}\widehat O_\alpha\widehat O_\beta
\bigl(\Gamma^{(\alpha)}_{ij}\Gamma^{(\beta)}_{kl}-\Gamma^{(\alpha)}_{il}\Gamma^{(\beta)}_{kj}\bigr).\tag{A.6}
\end{equation}
Assume all the Fierz relations hold:
\begin{equation}
\widehat O_\alpha\widehat O_\beta=\sum_{\gamma,\delta}C^{\alpha\beta}_{\gamma\delta}\widehat O_\gamma\widehat O_\delta,
\qquad C^{\alpha\beta}_{\gamma\delta}=\frac1{N^2}\Tr[\Gamma^{(\gamma)}\Gamma^{(\alpha)}\Gamma^{(\delta)}\Gamma^{(\beta)}].\tag{A.7}
\end{equation}
Apply them to the second term of (A.6):
\begin{align}
M_{ij;kl}=\frac1{N^2}\biggl[&\sum_{\alpha,\beta}\widehat O_\alpha\widehat O_\beta\Gamma^{(\alpha)}_{ij}\Gamma^{(\beta)}_{kl}\notag\\
-&\sum_{\alpha,\beta,\gamma,\delta}\widehat O_\gamma\widehat O_\delta C^{\alpha\beta}_{\gamma\delta}\Gamma^{(\alpha)}_{il}\Gamma^{(\beta)}_{kj}\biggr].\tag{A.8}
\end{align}
Cyclic invariance gives
\begin{equation}
C^{\alpha\beta}_{\gamma\delta}=\frac1{N^2}\Tr[\Gamma^{(\alpha)}\Gamma^{(\delta)}\Gamma^{(\beta)}\Gamma^{(\gamma)}]=C^{\delta\gamma}_{\alpha\beta}.\tag{A.9}
\end{equation}
The corresponding matrix Fierz relation then gives
\begin{align}
M_{ij;kl}=\frac1{N^2}\biggl[&\sum_{\alpha,\beta}\widehat O_\alpha\widehat O_\beta\Gamma^{(\alpha)}_{ij}\Gamma^{(\beta)}_{kl}\notag\\
-&\sum_{\gamma,\delta}\widehat O_\gamma\widehat O_\delta\Gamma^{(\delta)}_{ij}\Gamma^{(\gamma)}_{kl}\biggr]=0.\tag{A.10}
\end{align}
Here the matrix Fierz relations follow directly from completeness,
$\sum_\alpha\Gamma^{(\alpha)}_{ij}\Gamma^{(\alpha)}_{kl}=N\delta_{il}\delta_{jk}$.
All $2\times2$ minors vanish, so $\operatorname{Rank}(R)\leq1$. Since $R\ne0$, its rank is one.\hfill Q.E.D.

\section{Inductive proof of $R$-cycle relation for arbitrary $N$-cycles}
For the positive-semidefinite Hermitian matrix $R=bb^\dagger=\langle b|b\rangle\rho$, rank one is equivalent to the minor constraints (A.2). Anchor the induction with the two-cycle relation
\begin{equation}R_{i_1i_2}R_{i_2i_1}=R_{i_1i_1}R_{i_2i_2}=\prod_{\xi=1}^2R_{i_\xi i_\xi}.\tag{A.11}\end{equation}
Assume the following relation for an arbitrary $N$-cycle:
\begin{equation}R_{i_1i_2}R_{i_2i_3}\cdots R_{i_Ni_1}=\prod_{\xi=1}^N R_{i_\xi i_\xi}.\tag{A.12}\end{equation}
The inductive step $N\mapsto N+1$ is
\begin{align}
R_{i_1i_2}\cdots R_{i_Ni_{N+1}}R_{i_{N+1}i_1}
&=R_{i_1i_2}\cdots R_{i_{N-1}i_N}R_{i_Ni_1}R_{i_{N+1}i_{N+1}}\notag\\
&=\prod_{\xi=1}^{N+1}R_{i_\xi i_\xi}.\tag{A.13}
\end{align}
The first equality uses (A.2) and the second the induction hypothesis.\hfill Q.E.D.

Inserting $R=bb^\dagger$ cancels the amplitude moduli and leaves the phase-factor relation
\begin{equation}e^{i\phi_{i_1i_2}}e^{i\phi_{i_2i_3}}\cdots e^{i\phi_{i_Ni_1}}=1,\tag{A.14}\end{equation}
or
\begin{equation}\phi_{i_1i_2}+\phi_{i_2i_3}+\cdots+\phi_{i_Ni_1}=0\pmod{2\pi}.\tag{A.15}\end{equation}
Such consistency relations for relative phases $\phi_{ij}:=\phi_i-\phi_j$ (for $b_j=|b_j|e^{i\phi_j}$) are vital for the reconstruction procedures discussed in references~\cite{Chiang1997,Nakayama2019,Wunderlich2020,Wunderlich2021}. The graphs in the latter two references may be regarded as pictorial representations of consistency relations within $N$-cycles.

\end{document}